\documentclass[twoside,fleqn]{article}

\usepackage{espcrc2}
\usepackage{epsfig,psfig}
\usepackage{graphicx}
\usepackage[figuresright]{rotating}

\def\proof{\noindent{\sl Proof:}\kern0.6em}

\def\dual{\mathstrut^*\kern-0.1em}

\def\lvec#1{\setbox0=\hbox{$#1$}
    \setbox1=\hbox{$\scriptstyle\leftarrow$}
    #1\kern-\wd0\smash{
    \raise\ht0\hbox{$\raise1pt\hbox{$\scriptstyle\leftarrow$}$}}
    \kern-\wd1\kern\wd0}
\def\rvec#1{\setbox0=\hbox{$#1$}
    \setbox1=\hbox{$\scriptstyle\rightarrow$}
    #1\kern-\wd0\smash{
    \raise\ht0\hbox{$\raise1pt\hbox{$\scriptstyle\rightarrow$}$}}
    \kern-\wd1\kern\wd0}

\def\nabstar#1{\nabla\kern-0.5pt\smash{\raise 4.5pt\hbox{$\ast$}}
               \kern-4.5pt_{#1}}

\def\drvstar#1{\partial\kern-0.5pt\smash{\raise 4.5pt\hbox{$\ast$}}
               \kern-5.0pt_{#1}}

\def\MeV{{\rm MeV}}
\def\GeV{{\rm GeV}}

\def\fm{{\rm fm}}

\def\psibar{\overline{\psi}}

\def\rhoprime{\rho\kern1pt'}
\def\rhobar{\bar{\rho}}
\def\rhobarprime{\rhobar\kern1pt'}
\def\rhobartilde{\kern2pt\tilde{\kern-2pt\rhobar}}
\def\rhobartildeprime{\kern2pt\tilde{\kern-2pt\rhobar}\kern1pt'}

\def\zetabar{\bar{\zeta}}
\def\zetaprime{\zeta\kern1pt'}
\def\zetabarprime{\zetabar\kern1pt'}

\def\diracstar#1#2{
    \setbox0=\hbox{$\gamma$}\setbox1=\hbox{$\gamma_{#1}$}
    \gamma_{#1}\kern-\wd1\kern\wd0
    \smash{\raise4.5pt\hbox{$\scriptstyle#2$}}}
\def\momp#1#2{
    \setbox0=\hbox{${#1}$}\setbox1=\hbox{${#1}_{#2}$}
    {#1}_{#2}\kern-\wd1\kern\wd0
    \smash{\raise4.5pt\hbox{$\scriptscriptstyle +$}}}
\def\momm#1#2{
    \setbox0=\hbox{${#1}$}\setbox1=\hbox{${#1}_{#2}$}
    {#1}_{#2}\kern-\wd1\kern\wd0
    \smash{\raise4.5pt\hbox{$\scriptscriptstyle -$}}}
\def\mompm#1#2{
    \setbox0=\hbox{${#1}$}\setbox1=\hbox{${#1}_{#2}$}
    {#1}_{#2}\kern-\wd1\kern\wd0
    \smash{\raise4.5pt\hbox{$\scriptscriptstyle \pm$}}}
\def\smomp#1#2{
    \setbox0=\hbox{${#1}$}\setbox1=\hbox{${#1}_{#2}$}
    {#1}_{#2}\kern-\wd1\kern\wd0
    \smash{\raise3pt\hbox{$\scriptscriptstyle +$}}}
\def\smomm#1#2{
    \setbox0=\hbox{${#1}$}\setbox1=\hbox{${#1}_{#2}$}
    {#1}_{#2}\kern-\wd1\kern\wd0
    \smash{\raise3pt\hbox{$\scriptscriptstyle -$}}}
\def\smompm#1#2{
    \setbox0=\hbox{${#1}$}\setbox1=\hbox{${#1}_{#2}$}
    {#1}_{#2}\kern-\wd1\kern\wd0
    \smash{\raise3pt\hbox{$\scriptscriptstyle \pm$}}}

\def\csw{c_{\rm sw}}

\def\ba{b_{\rm A}}
\def\bp{b_{\rm P}}

\def\msbar{{\rm \overline{MS\kern-0.05em}\kern0.05em}}
\def\smallmsbar{\small\overline{\hbox{MS\kern-0.10em}}
                \hbox{\kern0.10em}}

\def\mbar{\overline{m}}

\def\mK{m_{\rm K}}
\def\mps{m_{\rm PS}}

\def\mqu{m_{\rm u}}
\def\mqd{m_{\rm d}}
\def\mqs{m_{\rm s}}
\def\mhat{\widehat{m}}

\def\fps{F_{\rm PS}}
\def\fK{F_{\rm K}}

\newcommand{\be}{\begin{equation}}
\newcommand{\ee}{\end{equation}}
\newcommand{\bea}{\begin{eqnarray}}
\newcommand{\eea}{\end{eqnarray}}
\newcommand{\bi}{\begin{itemize}}
\newcommand{\ei}{\end{itemize}}
\newcommand{\eq}[1]{eq.\,(\ref{#1})}

\newcommand{\nf}{N_{\rm f}}

\newcommand{\rb}[1]{\raisebox{1.5ex}[-1.5ex]{#1}}

\def\Leff{{\cal L}_{\sf eff}}
\def\msea{{m^{\rm sea}}}

\def\mref{m_{\sf ref}}
\def\RFG{R_{\rm M}}
\def\RF{R_{\rm F}}

\def\Tr{{\sf Tr}\,}

\def\gsim{{\mathrel{\raise2pt\hbox to 8pt{\raise -5pt\hbox{$\sim$}\hss{$>$}}}}}
\def\rsim{{\mathrel{\raise2pt\hbox to 8pt{\raise -5pt\hbox{$\sim$}\hss{$>$}}}}}
\def\lsim{{\mathrel{\raise2pt\hbox to 8pt{\raise -5pt\hbox{$\sim$}\hss{$<$}}}}}

\begin{document}

\title{
\vspace{-3.6cm}
       \begin{flushright}
       {\normalsize
        \tt DESY 02-164\\[-0.2cm]
            October 2002}\\[-0.2cm]
       \end{flushright}
       \vspace{1.0cm}
Chiral Effective Lagrangian and Quark Masses
\thanks{Plenary talk presented at {\it Lattice 2002}, MIT, Cambridge,
USA,24--29 June 2002.}
}

\author{Hartmut Wittig\address{%
        DESY, Theory Group, Notkestr. 85, D-22603 Hamburg, Germany}
}
      
\begin{abstract}
The status of lattice determinations of quark masses is reviewed (with
the exception of $m_{\rm b})$. Attempts to extract the low-energy
constants in the effective chiral Lagrangian are discussed,
with special emphasis on those couplings which are required to test the
hypothesis of a massless up-quark.
Furthermore, the issue of quenched chiral logarithms is addressed.
%
%possible cheating on space
\vspace{-0.2in}
\end{abstract}

\maketitle
\noindent

% more cheating
% \vspace{-0.1in}
\vspace{-0.3cm}
\section{INTRODUCTION}

Recently, many studies of QCD at low energies have investigated the
interplay between lattice simulations and Chiral Perturbation Theory
(ChPT). I shall start the review of these activities by recalling the
basic features of ChPT and point out what can be learned from
combining ChPT with lattice simulations.

Chiral Perturbation Theory is a systematic expansion in the 4-momentum
$p$ and the quark masses $\mqu,\,\mqd,\,\mqs$ about the massless limit
\cite{GaLeu_ChPT}. In the low-energy regime its information content is
equivalent to that of QCD, and this has been exploited in many
applications, e.g. in the calculation of pion scattering amplitudes
and quark mass ratios. Furthermore, ChPT provides valuable input for
lattice simulations. Here, the prediction of the quark mass dependence
of $\mps^2$ is surely the most widely used piece of information. In
addition, ChPT can model the volume dependence of observables, and
also relates different processes. For instance, the amplitudes of
$K\to\pi\pi$ decays can be expressed in terms of the theoretically
much simpler $K\to\pi$ transition.

However, Chiral Perturbation Theory is an effective,
non-renormalisable theory, parameterised in terms of a set of
empirical couplings, which are usually called ``low-energy constants''
(LECs). At lowest order the effective chiral Lagrangian reads
\bea
  \Leff^{(2)}=\frac{F_0^2}{2}\Big\{& &\hspace{-0.7cm}
    \frac{1}{2} \Tr\left(\partial_\mu 
    U^\dag\partial^\mu U\right) \nonumber\\
  & & +B_0\Tr\left({\cal M}(U+U^\dag) \right)\Big\},
\eea
where
\be
       U=\exp\left\{{i\,T^a\varphi^a}/{(\sqrt{2}F_0)}\right\}
\ee
$\mbox{\empty}$\\
is the field of Goldstone bosons, and ${\cal M}={\rm
diag}(\mqu,\mqd,\mqs)$ is the quark mass matrix. The LECs at leading
order are $B_0$ and $F_0$, where the latter corresponds to the pion
decay constant in the chiral limit.\footnote{Throughout this review I
use conventions where $F_\pi=93.3\,\MeV$.} At next-to-leading order
there are 12 more interaction terms and hence 12 new LECs,
$L_1,\ldots,L_{12}$.

The values of the LECs can be fixed by using experimental data and
comparing with the relevant expressions obtained in ChPT. It turns
out, however, that the complete set of LECs cannot be determined in
this way. In other words, the values of some of the LECs cannot be
fixed without resorting to additional theoretical assumptions. One
particular example is the value of $B_0$, which appears in the chiral
expansion of the pion mass at leading order:
\be
    m_\pi^2 = 2B_0\mhat,\quad \mhat=\textstyle\frac{1}{2}(\mqu+\mqd).
\ee
This shows that $B_0$ can only be determined from $m_\pi$
if the physical value of the quark mass $\mhat$ is known already. By
the same token, the quark mass $\mhat$ can only be inferred if an
estimate for $B_0$ is available. However, $B_0$ drops out in suitably
chosen ratios of $m_\pi^2,\,\mK^2,\ldots$. Thus, while ChPT enables us
to compute quark mass ratios, it fails to provide an absolute
normalisation of their masses. Another reason why the complete set of
LECs cannot be determined from chiral symmetry considerations alone is
the fact that the effective chiral Lagrangian beyond leading order is
invariant under a symmetry transformation involving the LECs and the
mass matrix ${\cal M}$. This is the famous ``Kaplan-Manohar
ambiguity'' \cite{KapMan_85}.

At this point it is clear that lattice simulations can in turn provide
valuable input for ChPT. By studying the quark mass dependence of
Goldstone bosons in a simulation one can determine the value of $B_0$,
or, equivalently provide an absolute normalisation of quark
masses. Since the Kaplan-Manohar (KM) ambiguity is not a symmetry of
QCD, it is possible to resolve it by determining the values of LECs
from an approach based on first principles. It must be kept in mind,
though, that a successful combination of lattice QCD and ChPT requires
sufficient overlap between the range of quark masses used in
simulations and the region of validity of the chiral expansion.  This
requirement is crucial if ChPT is used to extrapolate results from
simulations performed for relatively heavy quarks and the regime of
the physical $u$- and $d$-quark masses \cite{panel_lat02}.

% \P The
%ultimate goal would be to fix the LECs from lattice simulations of the
%underlying theory of QCD. ChPT could then be used to establish contact
%between lattice simulations performed for relatively heavy quarks and
%the regime of very light quark masses. This is of particular
%importance for partially quenched simulations, where the masses of
%valence quarks differ from those of the sea quarks. Such a strategy
%has been anticipated e.g. in ref.~\cite{ShaSho_00}. \P

The remainder of this review covers a summary of the current status of
lattice determinations of the light quark masses, quenched chiral
logarithms, and LECs at NLO.% in the chiral Lagrangian.

\vspace{-0.2cm}
\section{QUARK MASSES}

Recent years have seen a great deal of progress in lattice
determinations of quark masses. Some of the dominant systematic
effects are now under much better control, chiefly due to using
improved lattice actions and non-perturbative renormalisation, as
implemented either via the RI/MOM scheme \cite{MPSTV_94} or the
Schr\"odinger functional (SF) \cite{mbar:pap1}. Several collaborations
have started to quantify quenching errors, by performing simulations
with dynamical quarks. Finally, the results in the quenched
approximation are also being tested in simulations employing exact
chiral symmetry on the lattice.
Here we will focus on determinations of $\mqs$, $\mhat$ and the charm
quark mass. The mass of the $b$-quark is discussed in R. Sommer's
contribution to this conference \cite{Rainer_lat02}.  Earlier reviews
can be found in refs. \cite{Lubicz_lat00,HW_S2000,Kaneko_lat01}.

\vspace{-0.2cm}
\subsection{Wilson and staggered fermions}

The LEC $B_0$ is obtained straightforwardly by mapping out the quark
mass dependence of pseudoscalar meson masses. In order to compute,
say, $\mqs+\mhat$ one then has to specify the lattice scale and the
quantity which fixes the bare physical quark mass. For instance, using
$r_0$ to set the scale \cite{pot:r0} and the kaon mass $\mK$
(``$K$-input'') one has
\be
   \textstyle\frac{1}{2}(\mqs+\mhat) r_0 = Z\cdot
        \left.\displaystyle\frac{1}{{B_0 r_0}}\right|_{\mps=\mK}
        \times (\mK r_0)^2_{\rm exp}.
\label{eq_qmass_def}
\ee
Here, $Z$ denotes the renormalisation factor which relates the quark
mass in lattice regularisation to a reference continuum scheme
(e.g. $\msbar$), and $(\mK r_0)^2_{\rm exp}$ is the phenomenological
estimate of the kaon mass in units of the lattice scale. In order to
estimate $\mqs$, one either has to extract $\mhat$ separately through
an extrapolation to $(m_\pi r_0)^2$, or take the ratio $\mqs/\mhat$
from ChPT.

\begin{table*}[htb]
\caption{Results for $\mqs$ in the $\msbar$-scheme at
$\mu=2\,\GeV$ and for the ratio $\mqs/\mhat$, obtained using Wilson
(W) and staggered (KS) quarks ($K$-input). Where applicable, we
include information on the implementation of O($a$) improvement
(non-perturbative (NP), mean-field perturbative (MF), tree-level
(tree)), as well as the method for non-perturbative renormalisation
(RI or SF).}
\label{tab_mstrange}
\begin{center}
\begin{tabular}{l c c c c c c l l}
\hline\hline
Collaboration & Ref. & action & Impr. & $a$ [fm] & Scale & Ren.
 & $\mqs\,[\MeV]$ & $\mqs/\mhat$ \\
\hline
$\rm SPQ_{cd}R$ & \cite{quark:SPQR02}
 & W & NP & 0 & $m_{K^*}$ & RI & 106(2)(8) & 24.3(2)(6) \\
CP-PACS  & \cite{quen_CPPACS_02} & W  & & 0 & $m_\rho$ & & 114(2)$(^6_3)$
  & 26.5$(^{5.1}_{3.4})$ \\
CP-PACS  & \cite{dyn_CPPACS_00} & W/Iwas. & MF & 0 & $m_\rho$ &
 & 110$(^3_4)$  & 25.0(1.4) \\
APE & \cite{quark:APE_prop} & W & NP & 0.07 & $m_{K^*}$ & RI & 111(9)
 & 23.1(3.1) \\
QCDSF & \cite{quark:QCDSF_99} & W & NP &  0 & $r_0$ & SF & 105(4)
 & 23.9(1.4) \\
ALPHA/UKQCD & \cite{mbar:pap3} & W & NP & 0 & $\fK$ & SF &  ~\,97(4) &  \\
JLQCD & \cite{quark:JLQCD_stag99} & KS & & 0 & $m_\rho$ & RI & 106(7)
 & 25.1(2.4) \\
APE   & \cite{quark:APE_99} & W & NP & 0.07 & $m_{K^*}$ & RI & 111(12) &
24.7(3.4) \\
GGRT  & \cite{quark:GGRT_98} & W & tree & 0.07 & $m_{K^*}$ & RI & 130(2)(18)
 & 22.8(4.5) \\
\hline\hline
\end{tabular}
\end{center}
\vspace{-0.5cm}
\end{table*}

A compilation of recent results for $\mqs$ and the ratio $\mqs/\mhat$
is shown in Table \ref{tab_mstrange}. A marked feature is that
estimates for $\mqs$ in the continuum limit have stabilised, in
contrast to the situation seen a few years ago. The table also shows
that lattice estimates for the ratio $\mqs/\mhat$, which are all based
on chiral extrapolations, are broadly consistent with the result in
ChPT at NLO
\cite{Leutwyler_96}:
\be
    \mqs/\mhat = 24.4\pm1.4.
\ee
However, closer inspection reveals small but significant deviations
among the results for $\mqs$. Therefore, for the results to be
consistent, these deviations must be due to quenching effects, which
are caused by different choices for the lattice scale. We will now
attempt to convert some of the results in Table~\ref{tab_mstrange} to
a common scale, in order to check whether consistency is satisfied.

To this end we consider two lattice scales, $Q$ and $Q^\prime$. From
the definition of the quark mass in \eq{eq_qmass_def} it follows that
the strange quark mass $\mqs^{(Q)}$ in MeV, estimated using $Q$, is
related to its counterpart $\mqs^{(Q^\prime)}$ via
\be
   \mqs^{(Q)}\,[\MeV] = 
        \underbrace{
             \left(Q\over Q^\prime\right)_{\rm lat} 
             \left(Q^\prime\over Q\right)_{\rm exp} 
        }_{\textstyle F}
        \mqs^{(Q^\prime)}\,[\MeV].
\ee
Here, the subscripts ``lat'' and ``exp'' refer to lattice and
experimental estimates of the scale ratio, respectively. The deviation
of the factor $F$ from unity is indicative of the relative quenching
effects, when either $Q$ or $Q^\prime$ is chosen to set the scale. I
have taken results for vector meson masses from
refs. \cite{quark:SPQR02,quen_CPPACS_02,quark:JLQCD_stag99} and for
$\fK$ from \cite{mbar:pap3} to determine the ratio $(Q/Q^\prime)_{\rm
lat}$ in the {\it continuum limit} for $Q=r_0^{-1}$ and
$Q^\prime=m_\rho,\,m_{\rm K^*}, \fK$. The factor $F$ can then be
computed using the phenomenological values of $r_0=0.5\,\fm$ and
$Q^\prime$. The results are shown in Table~\ref{tab_QQprime}. The
first observation is that $F$ varies by up to 15\% for the various
scales that are compared.
% Furthermore, one obtains consistent
% estimates for $F$ if $m_\rho$ is computed with either Wilson or
% staggered quarks.
Once the results have been converted to the common scale $r_0$, the
estimates for $\mqs$ in the continuum limit show remarkable
consistency. This is actually surprising, since the various
simulations differ significantly, not only in terms of the fermionic
discretisation, but also by the chosen method to relate the lattice
estimate to the $\msbar$ scheme (mean-field perturbative,
non-perturbative via RI/MOM or SF). Lastly, I want to stress that a
meaningful comparison of this kind can only be made in the continuum
limit.

\begin{table}
\caption{Results for the conversion factor $F$ for various scales
$Q^\prime$ and fixed $Q=r_0^{-1}$ at zero lattice spacing. Also shown
are the estimates for $\mqs^{(Q)}$ expressed through the common scale
$r_0^{-1}$ in MeV ($\msbar$-scheme at $\mu=2\,\GeV$).}
\label{tab_QQprime}
%\begin{center}
\begin{tabular}{l r@{(}l c c l}
\hline\hline
Ref. & \multicolumn{2}{c}{$\mqs^{(Q^\prime)}$} & $Q^\prime$ & $F$
 & $\mqs^{(r_0)}$ \\
\hline
\cite{quark:JLQCD_stag99} & 106&7)  & $m_\rho$ & 0.90(1)  & {95(6)} \\ 
\cite{quen_CPPACS_02}     & 114&2)$(^6_3)$  & $m_\rho$ & 0.86(2) 
& {98(2)$(^6_3)$}\\ 
\cite{quark:SPQR02} & 106&2)(8) & $m_{K^*}$ & 0.87(3) & 92(2)(7) \\
\cite{mbar:pap3}    &  97&4) & {$\fK$}   & 1.02(2) & 99(4) \\
\hline\hline
\end{tabular}
%\end{center}
\vspace{-0.5cm}
\end{table}

\vspace{-0.3cm}
\subsection{Ginsparg-Wilson fermions}

Lattice actions that satisfy the Ginsparg-Wilson relation
\cite{GinsWil_82} and thus preserve chiral symmetry at non-zero
lattice spacing are now routinely used to compute many
phenomenologically interesting quantities. The most widely used
implementations are based on overlap \cite{overlap} or domain wall
fermions \cite{domainwall}. More recently, results for fixed-point
(FP) \cite{fixedpoint} or chirally improved (CI) \cite{chirimp}
actions, which both provide approximate solutions to the
Ginsparg-Wilson relation, have also become available (see
\cite{gatt_lat02} for a review).

The mass of the strange quark was actually one of the first quantities
computed using domain wall fermions \cite{BSW99}. Since then many
systematic effects have been studied: discretisation errors have been
estimated by comparing results at different lattice spacings, although
no continuum extrapolations have been performed so far. Quenching
effects have not been investigated, since the numerical effort to
simulate quenched Ginsparg-Wilson fermions is already comparable to a
dynamical simulation using Wilson quarks
\cite{panel_lat01}.

We have seen that non-perturbative renormalisation is an important
ingredient in order to enhance the credibility of lattice estimates
for quark masses. The RI/MOM prescription has already been applied for
both domain wall \cite{RBC_NPren} and overlap fermions
\cite{GHR_lat01}. The Schr\"odinger functional, due to its
inhomogeneous boundary conditions is somewhat harder to realise for
Ginsparg-Wilson fermions (an attempt has been made in
ref. \cite{CPPACS_SF_DWF} in the case of domain wall fermions). The
authors of \cite{HJLW_01} have proposed a general strategy to compute
the renormalisation factors of quark bilinears for overlap fermions in
the SF via an intermediate Wilson regularisation. This procedure
avoids the direct formulation of the SF for Ginsparg-Wilson fermions,
at the expense of sacrificing the ability to predict the quantity used
to match to the Wilson results in the intermediate step.

\begin{table*}[htb]
\caption{Results for $\mqs$ and the ratio $\mqs/\mhat$, obtained using
domain wall (DW) and overlap (Ov) quarks. Conventions are identical to
Table~\protect\ref{tab_mstrange}.}
\vspace{0.1cm}
\label{tab_mstrange_GW}
\begin{center}
\begin{tabular}{l c c r@{.}l c l l}
\hline\hline
Collaboration & Ref. & action & \multicolumn{2}{c}{$a$ [fm]} & NP-Ren.
 & $\mqs\,[\MeV]$ & $\mqs/\mhat$ \\
\hline
        &  & DW/DBW2 & $\sim$0&099  &   & 133(3) & \\
\rb{RBC}& \rb{\cite{quark:RBC_lat02}} & DW/W & $\sim$0&099 & \rb{RI}
       & 126(3) & \rb{$\sim26$} \\ \hline
CP-PACS & \cite{CPPACS_BK_DWF_01}& DW & $\ge$0&066  &        & ~\,99(2)(6)
        & 26.3(2.3) \\ \hline
        &           &         & 0&093  &        & 105(6)(21) & \\ 
\rb{RBC}& \rb{\cite{RBC_wingate_00}}
        & \rb{DW} & 0&123  &\rb{RI} & 100(5)(20) & \\ \hline
BSW     & \cite{BSW99} & DW      & $\ge$0&073 &    & ~\,96(26) & \\ \hline
        &              &         & 0&090      &    & 110(7) 
        & 24.41(5) \\
\rb{DeGrand} & \rb{\cite{DeGrand_lat02}} & \rb{Ov} & 0&125 & & 105(7)
        & 24.40(4) \\ \hline
C+H     & \cite{ChiuHsieh02_2}   & Ov    & 0&147    &         & 115(8)
        & 21.9(2) \\ \hline
GHR   & \cite{GHR_01}  & Ov    & 0&086       & {RI} & 102(6)(18) & \\ \hline
        &          &           & 0&093      &    & ~\,96(5)(7)
        & 26.1(3.0) \\
\rb{HJLW}& \rb{\cite{HJLW_lat02}}&\rb{Ov}& 0&123    &\rb{SF}  & ~\,99(4)(7)
         & 25.9(3.4) \\
\hline\hline
\end{tabular}
\end{center}
\vspace{-0.5cm}
\end{table*}

%\begin{figure}[htb]
%\begin{center}
%%\hspace{-2pt}
%\psfig{file=mstrange_GW.ps,width=9cm}
%\vspace{-1.cm}
%\caption{Scaling plot of the results for $\mqs$ listed in
%Table~\protect\ref{tab_mstrange_GW}. Also the comparison with Wilson
%results is shown (full circles).}
%\label{fig_quark_GW}
%\end{center}
%\vspace{-1.0cm}
%\end{figure}

Results for $\mqs$ and $\mqs/\mhat$ obtained using either domain
wall or overlap fermions are compiled in Table~\ref{tab_mstrange_GW}.
The numbers for $\mqs$ are broadly consistent among each other, and
also with the continuum estimates discussed earlier. However, the
overall uncertainties are still somewhat larger compared with Wilson
or staggered fermions. In most cases, the quoted systematic
uncertainties are dominated by lattice artefacts, which have not been
removed by performing continuum extrapolations. However, the overall
dependence of $\mqs$ on the lattice spacing appears to be fairly
weak. These indications of good scaling properties of Ginsparg-Wilson
fermions must be corroborated with more statistics and systematic
investigations of renormalisation and finite-$a$ effects.

\vspace{-0.2cm}
\subsection{Dynamical simulations}

\begin{figure}[htb]
\begin{center}
\vspace{-1.0cm}
\psfig{file=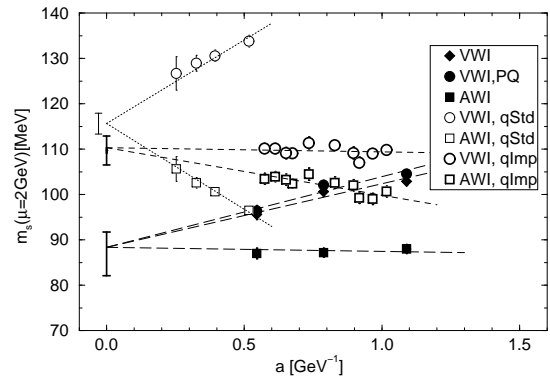,width=8cm}
\vspace{-1.8cm}
\caption{Scaling plot of dynamical and quenched estimates for $\mqs$
by CP-PACS \protect\cite{dyn_CPPACS_00}.}
\label{fig_cppacs_quark_dyn}
\vspace{-0.7cm}
\end{center}
\end{figure}

The most important challenge for current lattice studies is surely the
quantification of quenching effects. Perhaps the most comprehensive
study so far has been published by CP-PACS some time ago
\cite{dyn_CPPACS_00}. They investigated $\nf=2$ flavours of dynamical
quarks, using mean-field improved Wilson fermions and the Iwasaki
gauge action at three different values of the lattice spacing in the
range $a\approx0.1-0.2\,\fm$. Input values for the dynamical quark
masses were chosen such that $\mps/m_{\rm V}=0.58-0.8$. Estimates of
$\mqs$ based either on vector or axial vector Ward identities (VWI and
AWI respectively) can be extrapolated to a common value in the
continuum limit. By comparing their results to quenched data, CP-PACS
find that dynamical simulations lead to a decrease in the value of
$\mqs^\msbar$ by about 25\%. The dynamical data also appear to be less
sensitive to the choice of quantity used to fix the bare strange quark
mass ($\mK$ or $m_\phi$). The main findings of CP-PACS are summarised
in Fig.~\ref{fig_cppacs_quark_dyn}, as well as in the estimates
\be
   \mqs^\msbar(2\,\GeV)=88(^4_6)\,\MeV,\quad \mqs/\mhat\approx26.
\ee

While CP-PACS's results represent a major step forward, the
limitations of current dynamical simulations are apparent. In
particular, smaller dynamical quark masses should be simulated in
order to quantify quenching effects more reliably. Moreover,
simulating two flavours of dynamical quarks still does not represent
the physical situation. Discretisation errors must be
investigated more closely, by simulating smaller lattice spacing and
comparisons with alternative fermionic discretisations. Finally,
non-perturbative renormalisation should be implemented. Attempts to
improve the situation in each of these areas have been reported at
this conference.

The JLQCD and UKQCD Collaborations have both used non-perturbatively
O($a$)-improved Wilson fermions for $\nf=2$ \cite{JanSom_97} at
$\beta=5.2$. Whereas JLQCD \cite{Kaneko_lat02} uses sea quark masses
corresponding to $\mps/m_{\rm V}=0.6-0.8$, UKQCD simulates even
lighter masses, thereby running the risk of suffering from
finite-volume effects. JLQCD have reported finite-size effects at the
$3-5\%$ level only at their lightest quark mass. They also extrapolate
their results in $\msea$ to the physical $u$ and $d$-quark masses. The
quark masses are defined through the PCAC relation (AWI), and they
expect some systematic effects to cancel, since the AWI mass is
defined through a ratio of matrix elements. The results for $\mqs$ and
$\mhat$ are $20-30\%$ smaller than in the quenched
approximation. Scaling violations appear to be small, as there is good
agreement with the previous CP-PACS results \cite{dyn_CPPACS_00}, but
a continuum extrapolation is still lacking.

UKQCD perform extrapolations in the valence quark mass at fixed values
of $\msea$ and then monitor the results for decreasing sea quark mass,
in order to search for trends which indicate the presence of dynamical
quark effects. Unlike JLQCD they define the quark mass via the vector
Ward identity. UKQCD's results show no clear signs of dynamical quark
effects: although there is some trend in the data as $\msea$ is
decreased, it is statistically not significant, and the same is true
for the comparison with quenched data at a matched value of the
lattice spacing. Their result for the ratio $\mqs/\mhat$ is consistent
with ChPT.

Hein et al. \cite{Hein_lat02} have used dynamical configurations
provided by the MILC Collaboration \cite{MILC_nf3_01} for 2+1 flavours
of improved staggered quarks. In this formulation the large
flavour-changing interactions, which are typically encountered for
conventional staggered quarks are suppressed due to the use of fat
links. The lightest quark mass in \cite{MILC_nf3_01} corresponds to
$\mps/m_{\rm V}\approx0.37$, at a lattice spacing of
$a\approx0.13\,\fm$. Mean-field improved perturbation theory was used
when matching to the $\msbar$-scheme; the one-loop coefficients were
found to be of the same order of magnitude as for Wilson
fermions. Results for $\mqs+\mhat$ obtained for $\nf=0,\,2$ and 2+1
dynamical flavours show that the large dependence on the quantity that
sets the lattice scale is greatly reduced in the dynamical case. Low
values for the light quark masses are typically preferred; for
$\nf=2+1$ flavours Hein et al. quote
\be
    (\mqs+\mhat)^\msbar(2\,\GeV) = 78\pm14\,\MeV,
\ee
where the scale has been taken from the $K^*-K$ mass splitting. The
quoted error includes the uncertainty from neglecting the two-loop
term in the perturbative matching procedure, which is estimated to be
as large as 20\%.  The value of $(\mqs+\mhat)$ in the dynamical case
is lower by 15 -- 20\% compared with the corresponding quenched
result. For a better understanding of the systematics in the improved
staggered formulation, it would, in my opinion, be extremely helpful
if a quenched value for $\mqs$ were available in the continuum limit.

\begin{figure}[htb]
%\begin{center}
\vspace{-0.8cm}
\psfig{file=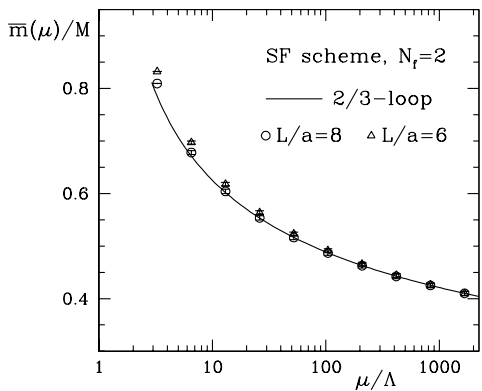,width=7cm}
\vspace{-1.cm}
\caption{Running quark mass in the SF scheme for $\nf=2$
\protect\cite{knechtli_lat02}.}
\label{fig_mbar_nf2}
\vspace{-0.7cm}
%\end{center}
\end{figure}

The ALPHA Collaboration has reported results on the non-perturbative
running of the quark mass in the SF scheme for the 2-flavour case
\cite{knechtli_lat02}. This is one part of the two-step procedure,
which will ultimately relate the bare current quark mass on the
lattice to the renormalisation group invariant (RGI) quark mass $M$
\cite{mbar:pap1}. In Fig.~\ref{fig_mbar_nf2} the running mass in the
SF scheme is plotted as a function of the renormalisation scale. One
sees that the perturbative evolution follows the numerical data down
to fairly small values of the scale. At its lowest value, i.e. at
$\mu=1/(2L_{\rm max})$, ALPHA obtain their preliminary result for the
matching of the running mass in the SF scheme to the RGI quark mass:
\be
    \frac{M}{\mbar_{\rm SF}}=1.236(15),\quad \ln(\Lambda
    L_{\rm max}) = -1.85(13).
\ee
The scale $L_{\rm max}$ must still be related to some physical
quantity. Further studies will include data at larger values of $L/a$,
so that lattice artefacts can be eliminated completely.

\vspace{-0.3cm}
\subsection{The charm quark mass}

The charm quark is too heavy to be described by ChPT, and too light
for an efficient non-relativistic treatment. In two recent projects
\cite{SPQCD_charm01,ALPHA_charm} the charm quark mass was
computed using the relativistic formulation. In this case, the issue
of controlling lattice artefacts is even more important as in the
light quark sector. The ALPHA Collaboration \cite{ALPHA_charm} have
used O($a$) improved Wilson fermions in the quenched approximation at
four $\beta$-values. They employ several different definitions of the
RGI charm quark mass, e.g.
\bea
  & & r_0M_{\rm c} = Z_{\rm M}\,r_0m_{\rm c}
       \left[ 1+(\ba-\bp)am_{\rm q,c}\right] \\
  & & r_0M_{\rm c} = Z_{\rm M}Z\,
       r_0m_{\rm q,c}\left[ 1+b_{\rm m}am_{\rm q,c}\right],
\eea
where $m_{\rm c}$ and $m_{\rm q,c}$ denote the bare masses defined
through the axial and vector Ward identities, respectively.
Non-perturbative values of renormalisation factors and improvement
coefficients as reported in \cite{mbar:pap1,ALPHA_babp_00} have been
used throughout, in order to guarantee a controlled extrapolation to
the continuum limit. The input quantities which fix the bare charm
quark mass in terms of meson masses were $m_{\rm D_s},\,\mK$ and the
ratio $\mqs/\mhat$ from ChPT. In this way chiral extrapolations can be
avoided completely. By demanding that different definitions of the RGI
charm quark mass extrapolate to a common continuum limit, they find
$r_0M_{\rm c}=4.19(11)$ at zero lattice spacing, which translates into
\be
    \mbar^\msbar_{\rm c}(\mbar_{\rm c})=1.301(34)\,\GeV,
\ee
where the 4-loop RG functions in the $\msbar$ scheme have been
used. This result agrees well with the estimate by the $\rm SPQ_{cd}R$
Collaboration \cite{SPQCD_charm01}
\be
    \mbar^\msbar_{\rm c}(\mbar_{\rm c})=1.26(4)(12)\,\GeV,
\ee
which has been obtained at $a\approx0.07\,\fm$.

\vspace{-0.2cm}
\subsection{Summary: Quark masses}

A detailed comparison of quenched data shows that the mass of the
strange quark is probably the most precisely determined quantity in
the quenched approximation: for a given choice of lattice scale, the
relative accuracy in the continuum limit is about 5\%. If the
$K$-meson is used to fix the bare $\mqs$, the final result varies
between 95 and 115 $\MeV$, depending on the choice of lattice
scale. There is a 20\% increase in $\mqs$ if the mass is fixed through
a vector meson like the $\phi$.

It is now clear that current lattice estimates for the strange quark
mass are completely dominated by quenching effects. First attempts to
quantify the quenching error indicate a decrease in $\mqs$ of about
20\%, but simulations at smaller quark masses and lattice spacings are
required to confirm this.

Estimates for the ratio $\mqs/\mhat$ based on chiral extrapolations
tend to agree with ChPT, both for quenched and unquenched
simulations. 

Finally, non-perturbative renormalisation and O($a$) improvement
enable one to perform controlled continuum extrapolations of the charm
quark mass, thereby yielding precise quenched estimates.

\vspace{-0.2cm}
\section{QUENCHED\,CHIRAL\,LOGARITHMS}

In this section we will be concerned with quenching artefacts in the
chiral expansion. It is well known that the quark mass behaviour of
hadron masses and matrix elements is modified in the quenched
approximation \cite{BerGol_qcl_92,sharpe_qcl_92,GolPal_qcl_97}. The
main qualitative difference is that flavour singlets do not decouple
in the quenched version of the effective low-energy theory. This is
because most of the loop contributions which lead to the decoupling of
flavour-singlet fields are removed, and hence the latter have to be
treated on the same footing as the octet fields. Incorporating the
flavour-singlet explicitly into the chiral Lagrangian introduces new
LECs: one is the mass scale $m_0$ of the flavour singlet, the other is
$\alpha_\Phi$, which multiplies the kinetic term in the
flavour-singlet part of the effective Lagrangian:
\be
      {\cal L}_{\rm sing}\propto{\alpha_\Phi}D_\mu\Phi_0
      D^\mu\Phi_0-{{m_0^2}} \Phi_0^2
\ee
The flavour-singlet propagator develops a double pole in the quenched
theory, which is proportional to $m_0^2$. Since $m_0$ does not vanish
in the chiral limit, this gives rise to a new type of chiral
logarithm, which diverges as the octet mass vanishes. The modified
chiral expansion for the pseudoscalar octet mass at NLO thus reads
\bea
     {\mps^2\over{2m}} &=& B_0\big\{
     1-\left({\delta}-\textstyle\frac{2}{3}{\alpha_\Phi}y\right)
     \left(\ln{y}+1\right) \nonumber \\
     & & +\left[\left(2\alpha_8-\alpha_5\right) 
                     -\textstyle\frac{1}{3}{\alpha_\Phi}\right]y \big\}.
\label{eq_mps2over2m}
\eea
For notational convenience we have introduced
\be
   \delta=\frac{m_0^2}{(4\pi F_0)^2\nf},\quad 
   y=\frac{2B_0m}{(4\pi F_0)^2}
\label{eq_ydef}
\ee
and rescaled the LECs $L_i$ to $\alpha_i=8(4\pi)^2 L_i$. A
phenomenological estimate of $\delta$ is obtained from the relation
\be
   m_0^2=m_{\eta^\prime}^2+m_\eta^2-2\mK^2
\label{eq_m0sq}
\ee
which gives $\delta\approx0.18$. There are several methods which allow
the determination of $\delta$ and $\alpha_\Phi$. The most
straightforward is to study the quark mass dependence of $\mps^2$ and
compare it with \eq{eq_mps2over2m}. Another method is to compute the
ratio of flavour-singlet and flavour-octet contributions to the
quenched $\eta^\prime$ correlator. The former are the so-called
``hairpin diagrams''. Finally, the Witten-Veneziano formula
\cite{WitVen_79} provides a link between the flavour-singlet mass
$m_0$ and the topological susceptibility $\chi_{\rm t}$:
\be
     m_0^2={2\nf\chi_{\rm t}\over{F_\pi^2}}.
\ee
The effects of quenched chiral logarithms are expected to show up very
near the chiral limit. However, in this regime quenched QCD is
afflicted with the occurrence of so-called ``exceptional
configurations'', i.e. unphysical near-zero modes, if conventional
discretisations such as Wilson fermions are used. Recently this has
stimulated the use of discretisations which do not suffer from this
problem, such as overlap and domain wall fermions, fixed point actions
and twisted mass QCD (tmQCD) \cite{tmQCD}. Also the ``Modified
Quenched Approximation'' \cite{MQA} in conjunction with Wilson
fermions has been applied. In addition, one has to avoid finite-volume
effects when going very near the chiral limit, which may fake the
signature of the quenched chiral logs. It is also clear that there is
a special r\^ole for discretisations which preserve chiral symmetry,
for which the Witten-Veneziano formula is exact
\cite{GiuRoTeVe_02,DeG_Hell_02}. Moreover, the comparison of
lattice data with the predictions of quenched lattice QCD at non-zero
lattice spacing is justified in this case.

\begin{table*}
\caption{Recent determinations of $\delta$. The second and third
columns show the minimum (bare) quark mass in $\MeV$ and the smallest
pion mass in units of the spatial lattice size, respectively}
\label{tab_qcl}
\vspace{0.1cm}
\begin{center}
\begin{tabular}{l c c c c c l}
\hline\hline
Collaboration & Ref.  & $m_{\rm q}$ [MeV] & $m_\pi^{\rm min}\,L$ & $a$\,[fm] &
action & $\delta$ \\
\hline
CP-PACS  & \cite{quen_CPPACS_02} & 20 & 6   & $0.05-0.1$ & W  & 0.10(2)   \\
QCDSF    & \cite{QCDSF_lat99}    & 22 & 4.7 & $0.05-0.1$ & SW & 0.14(2)   \\
FNAL     & \cite{MQA}            & 14 & 3   & 0.17       & SW & 0.065(13) \\
MILC     & \cite{MILC_nf3_01}    & 15 & 3.4 & 0.13       & KS & 0.061(3)  \\
Kentucky & \cite{Kent01}         & 19 & 3.2 & 0.157      & Ov & $0.2-0.4$ \\
RBC      & \cite{RBC_qcl_DWF}    & 28 & 3.8 & 0.104      & DW & 0.05(2)   \\
DeGrand/Heller & \cite{DeG_Hell_02} & 30 &  & 0.13       & Ov & 0.093(28) \\
Chiu+Hsieh & \cite{ChiuHsieh02_1}& 80 & 2.5 & 0.147      & Ov & 0.203(14) \\
Bern  & \cite{Bern_FPsim_02} & 38 & 2.6 & 0.13 & FP & $0.23(7)-0.30(18)$ \\
Kentucky & \cite{Kent_lat02} & 14 & 2.9 & 0.20 & Ov &  0.26(3) \\[0.1cm]
         &                   &    & 2.7 & 0.16 & FP &  0.17(2) \\
\rb{BGR} & \rb{\cite{BGR_lat02}}& & 2.9 & 0.15 & CI &  0.18(3) \\
\hline\hline
\end{tabular}
\end{center}
\vspace{-0.8cm}
\end{table*}

Lattice data for $\mps^2$ show very low sensitivity on $\alpha_\Phi$
in general. The RBC Collaboration finds \cite{RBC_qcl_DWF} that in the
studied mass range the combination $\alpha_\Phi\mps^2\ln\mps^2$ (or
$\alpha_\Phi y\ln{y}$ in \eq{eq_mps2over2m}) stays practically
constant, so that there is hardly any sensitivity to the value of
$\alpha_\Phi$. CP-PACS report \cite{quen_CPPACS_02} that fits in which
$\alpha_\Phi$ is treated as a free parameter are either unstable or
yield results that are consistent with zero. Thus, in most studies it
is assumed that $\alpha_\Phi=0$.

%\begin{figure}[htb]
%\begin{center}
%\vspace{-1.1cm}
%\psfig{file=delta.ps,width=7cm}
%\vspace{-1.0cm}
%\caption{Recent results for $\delta$. The vertical dashed line denotes
%the estimate $\delta=\approx0.18$ based on \protect\eq{eq_m0sq}.}
%\label{fig_delta}
%\vspace{-0.7cm}
%\end{center}
%\end{figure}

Table~\ref{tab_qcl} shows a compilation of recent determinations of
$\delta$. Overall one observes large variations in the current
estimates for $\delta$, with results from simulations employing
conventional fermionic discretisations being somewhat lower that those
using Ginsparg-Wilson fermions (exceptions are the low values quoted
in refs. \cite{RBC_qcl_DWF,DeG_Hell_02}). Despite the fact that the
sensitivity to quenched chiral logs is still quite low, there is
mounting evidence for a non-zero value for $\delta$. In fact, more
recent simulations, which employ lattice chiral symmetry, produce
results that are compatible with the phenomenological estimate based
on eq. (\ref{eq_m0sq}). However, a remaining matter of concern is the
possibility that finite-volume effects may distort the mass dependence
towards the chiral regime. This must be addressed in simulations on
larger volumes. So far no convincing signal for $\alpha_\Phi$ has been
observed, mainly because the sensitivity of lattice data is even worse
than for $\delta$.

\vspace{-0.2cm}
\section{IS THE UP-QUARK MASSLESS?}

Let us now return to the Kaplan-Manohar ambiguity. The fact that the
effective chiral Lagrangian beyond leading order is invariant under
simultaneous transformations of the mass matrix and a subset of LECs
implies an uncertainty in the size of the NLO correction to the mass
ratio $\mqu/\mqd$. This correction, $\Delta_{\rm M}$, is given by
\be
   \Delta_{\rm M} = \frac{(\mK^2-m_\pi^2)}{(4\pi F_\pi)^2}
   (2\alpha_8-\alpha_5) + \hbox{chiral logs}.
\ee
While $\alpha_5$ can be estimated from the ratio $\fK/F_\pi$, there exists
no experimental information which would allow an unambiguous
determination of $\alpha_8$ or indeed the linear combination
$2\alpha_8-\alpha_5$ \cite{Leut90,CoKapNel99}. The value of $\alpha_8$
and thus $\Delta_{\rm M}$ must then be fixed by invoking (plausible)
assumptions beyond chiral symmetry considerations. Proceeding in this
way Leutwyler \cite{Leutwyler_96} obtained
\be
 0<\Delta_{\rm M}\leq0.13.
\ee
This result excludes the possibility that $\mqu/\mqd=0$, which is only
possible for a large negative value for $\Delta_{\rm M}$. Therefore, a
massless up-quark, which presents a simple and elegant solution to the
strong CP problem is an unlikely scenario. However, an estimate of
$\Delta_{\rm M}$ based on first principles is still lacking. Given the
importance of the strong CP problem, it is desirable to tackle this
question by means of lattice determinations of $2\alpha_8-\alpha_5$.

 Here, partially quenched simulations of lattice QCD play an important
 r\^ole: for unequal sea and valence quark masses the effective chiral
 Lagrangian is parameterised in terms of the same LECs as the physical
 theory. Thus, as long as the correct number of dynamical quark
 flavours (i.e. $\nf=3$) is used, simulations based on unphysical mass
 combinations provide phenomenological information~\cite{ShaSho_00},
 provided that the regime of validity of ChPT and the range of
 simulated masses overlap.

A general method which allows for the extraction of LECs with good
statistical accuracy has been introduced by
ALPHA\,\cite{mbar:pap4}. Since the expressions of ChPT are valid for
arbitrary quark mass, one can introduce a reference value $\mref$
and define the ratios
\bea
  && \RF(x)  = \fps(m)/\fps(\mref)  \\
  && \RFG(x) = \left(\frac{2m}{\mps^2(m)}\right)\left/
               \left(\frac{2\mref}{\mps^2(\mref)}\right)\right.,
\eea
where $x$ is $m/\mref$. The ratios $\RF,\,\RFG$ are universal
functions of the dimensionless mass parameter $x$. All renormalisation
factors drop out, so that these quantities can be extrapolated to the
continuum limit straightforwardly. Due to the correlations between the
numerator and denominator one can expect good statistical precision
in numerical evaluations. The LECs at NLO can then be determined by
comparing numerical data for $\RF,\,\RFG$ with the corresponding
expressions in quenched
\cite{BerGol_qcl_92,sharpe_qcl_92,GolPal_qcl_97} or partially quenched
\cite{sharpe_pq_97,GolLeung_97} ChPT.

\begin{figure}[htb]
\begin{center}
\vspace{-0.6cm}
\psfig{file=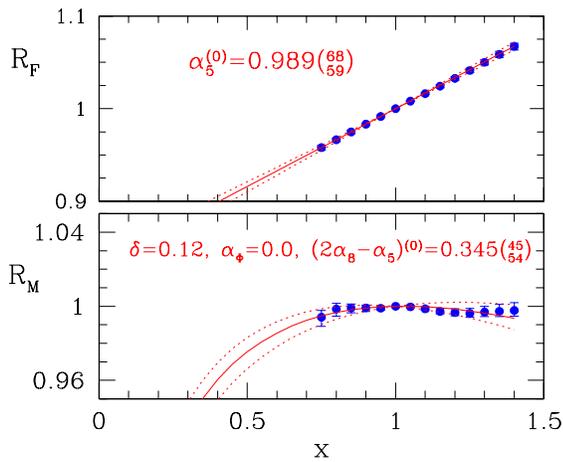,width=7.7cm}
\vspace{-1.2cm}
\caption{Results for $\RF$ and $\RFG$ in the quenched approximation
\protect\cite{mbar:pap4}.} 
\label{fig_RF_RFG_quen}
\vspace{-0.9cm}
\end{center}
\end{figure}

ALPHA have tested their method in the quenched approximation for
$\mref\approx\mqs$, and $x$ in the range $0.75\leq{x}\leq1.4$
\cite{mbar:pap4}. Their results are shown in
Fig. \ref{fig_RF_RFG_quen}. The ratio $\RF$, which is predicted by
quenched ChPT to rise linearly with the quark mass, is modelled very
well by the data, resulting in stable estimates for
$\alpha_5$. Numerical data for the ratio $\RFG$, however, show an
almost constant behaviour, while quenched ChPT predicts the presence
of linear terms, as well as chiral logs. Thus, if the range of
simulated masses lies indeed within the regime where ChPT at NLO is
valid, then the constant behaviour of the data must be the result of a
strong cancellation between the various contributions at NLO.

Bardeen et al. \cite{MQA} have performed global fits to pseudoscalar
correlation functions to extract masses and various LECs. Using the
pole shifting prescription of the ``Modified Quenched
Approximation'' (MQA) at $\beta=5.7$ and mean-field improved Wilson
fermions, they report an estimate for $\alpha_5$ which is about 3
times larger than ALPHA's. A large fraction of this discrepancy can be
attributed to the fact that renormalisation factors for the axial
current have not been taken into account \cite{Eichten_priv}. In fact,
if the data for the decay constant published in \cite{MQA} are
subjected to ALPHA's ratio method, then the estimates for $\alpha_5$
are consistent among the two studies. Thus, the source for the
original discrepancy is a combination of renormalisation effects and
lattice artefacts, with the ratio method being a lot more stable
against these systematic effects.

\begin{table*}
\caption{Results for the LECs at NLO for several values of $\nf$. The
last two lines display the results for the ``standard''
phenomenological estimates, as well as the numbers required for the
up-quark to be massless.}
\label{tab_LECs_NLO}
\vspace{0.1cm}
\begin{center}
\begin{tabular}{l c c l l l l}
\hline\hline
Collab. & Ref. & $\nf$ & $\quad\alpha_5$ & $\qquad\alpha_8$
 & $2\alpha_8-\alpha_5$ & comments \\
\hline
  &  &  0   & 0.99(6)(20)  &
 $\phantom{-}$0.50(4)(20)  & 0.35(5)(15)  &  $\delta=0.12$, $\alpha_\Phi=0$ \\
\rb{ALPHA}  & \rb{\cite{mbar:pap4}}  
 & ``3'' & 0.75(6)(20)  & $\phantom{-}$0.45(4)(20) & 0.15(5)(15) & \\[0.2cm]  
  &  &  2   & 1.22(13)(25) &
 $\phantom{-}$0.79(6)(21) & 0.36(10)(22) & \\ 
\rb{UKQCD} & \rb{\cite{ukqcd:mup0}}
 & ``3'' & 0.98(13)(24) & $\phantom{-}$0.59(6)(21) & 0.20(11)(22)
         & \\[0.2cm] 
OSU    & \cite{OSU} &  3   &              &             & 0.33(2)(17)
 & \\[0.2cm] 
 & \cite{BEG_94} & 3& 0.5(6)       & $\phantom{-}$0.76(4) &
 & ``standard'' \\ 
 & \cite{CoKapNel99}  & 3& 0.5(6)       & $-$0.9(4)   &
 & $\mqu=0$ \\ 
\hline\hline
\end{tabular}
\end{center}
\vspace{-0.5cm}
\end{table*}

The ratio method has also been applied in two recent simulations of
partially quenched QCD. UKQCD \cite{ukqcd:mup0} have simulated two
flavours of O($a$) improved dynamical Wilson fermions at $a=0.1\,\fm$
and a fixed value of the sea quark mass of $\msea\approx0.7\mqs$,
which corresponds to $\mps/m_{\rm{V}}\approx0.58$ \cite{csw202}. The
mass dependence of the ratios $\RF$ and $\RFG$ was subsequently
studied as a function of the valence quark mass at fixed $\msea$. The
overall behaviour of $\RF$ and $\RFG$ was found to be similar to what
is observed in the quenched case. However, the strict linearity of
$\RF$ for $\nf=0$ is modified by a slight curvature, which may signal
the expected presence of chiral logs. UKQCD's results for the LECs are
listed in Table \ref{tab_LECs_NLO}. 

The Ohio group \cite{OSU} have reported results for LECs from
simulations using $\nf=3$ flavours of dynamical staggered quarks. The
sea quarks are somewhat lighter than in UKQCD's study, while the
lattice spacings are larger ($a\approx0.15$ and $0.28\,\fm$). The
dominant systematic effects are large flavour-symmetry violations,
which are estimated by applying hypercubic blocking to the gauge
configurations before the computation of observables. The value of
$\alpha_5$ differs significantly when it is evaluated on blocked and
unblocked configurations. In the future this issue will be addressed
by comparing Monte Carlo data to ChPT for which the effects of
flavour-symmetry breaking are included \cite{cb_taste}.

From the compilation of results in Table \ref{tab_LECs_NLO} one
concludes that the $\nf$-dependence of LECs is quite weak.  If one
uses the expressions of ChPT for $\nf=3$, even though actually $\nf=0$
or~2 was used to generate the data, one still observes consistency
within errors (cases labelled $\nf=$``3'' in
Table~\ref{tab_LECs_NLO}). A comparison with the ``standard'' values
for the LECs estimated in the continuum, as well as with those values
that would be required to support the notion of a massless up-quark,
shows that the latter scenario is strongly disfavoured by lattice
calculations: the quark mass behaviour of $\RFG$ would have to be
radically different to accommodate a large negative value of
$\alpha_8$. In short, calculations based on first principles support
the analysis of ref. \cite{Leutwyler_96}.

The most important issue at this stage is whether or not the quarks
used in the simulations are light enough to justify the comparison
with ChPT. This has been addressed in several contributions to the
panel discussion on chiral extrapolations at this conference
\cite{panel_lat02}. Although lattice estimates for some of the LECs
make sense phenomenologically (e.g. UKQCD's result for $\alpha_5$ in
\cite{ukqcd:mup0} is consistent with the experimental value of
$\fK/F_\pi$ \cite{ukqcd:mup0}), there is evidence that the dependence
of $\mps^2$ and $\fps$ on the {\it sea} quark mass is not in agreement
with ChPT if the sea quark mass corresponds to $\mps=550 - 1000\,\MeV$
\cite{shoji_lat02,panel_lat02}. This affects attempts to perform
extrapolations in $\msea$ to the physical point defined by
$\mps/m_{\rm{V}}\approx0.17$. A similar study \cite{duerr_chpt_02}
concluded that sea quark masses of current simulations and ChPT at NLO
overlap only marginally.

In addition to many numerical results, there are also new analytic
developments. Aubin et al. \cite{cb_taste} have analysed the
structure of chiral logarithms under staggered flavour-symmetry
breaking for $\nf=2+1$ flavours. The basic idea is to generalise the
Lagrangian of Lee and Sharpe \cite{LeeSha}, which describes a single
staggered field up to O($a^2$), to the 3-flavour case. Normally, each
staggered flavour describes four internal fermions (called ``tastes''
in \cite{cb_taste}). In order to obtain one taste per flavour one then
takes the fourth root of the determinant. For the actual calculation
of chiral logs in the 2+1 flavour case, this implies that one starts
with 4+4 ``tastes'' and then cancels the unwanted loop contributions
by multiplying them with simple weight factors. The resulting
expressions for pseudoscalar meson masses fit the data by MILC
\cite{MILC_nf3_01} very well, unlike the corresponding formulae in
``ordinary'' partially quenched ChPT \cite{sharpe_pq_97,GolLeung_97}.

In a recent paper \cite{RupSho_01} Rupak and Shoresh have incorporated
the explicit chiral symmetry breaking of Wilson fermions directly into
the effective Lagrangian. The starting point is Symanzik's effective
action. Including all operators up to dimension 5, they find that
chiral symmetry is broken not only by the quark density $\psibar\psi$,
but also by the Pauli term $\psibar\sigma_{\mu\nu}F_{\mu\nu}\psi$,
which is multiplied by the coefficient $\csw$. Rupak and Shoresh then
consider simultaneous chiral expansions in the parameters
\be
    \epsilon\sim\frac{2B_0m}{\Lambda_\chi^2},\quad
    \delta\sim\frac{2W_0a\csw}{\Lambda_\chi^2}.
%    \epsilon\sim{2B_0m}/{\Lambda_\chi^2},\quad
%    \delta\sim{2W_0a\csw}/{\Lambda_\chi^2}.
\ee
The Pauli term generates additional operators in the effective chiral
Lagrangian, starting at leading order. Consequently, there are a
number of additional LECs. Up to NLO these are $W_0, W_4,\ldots, W_8$,
where the subscripts are chosen in analogy with the operators
appearing in the conventional Lagrangian. The resulting expressions
for pseudoscalar masses and decay constants in the partially quenched
case may be used to extract the LECs at small but non-zero lattice
spacing.

\vspace{-0.2cm}
\section{SUMMARY}

Lattice determinations of quark masses have entered a mature stage,
where the dominant systematic error is quenching. The quantification
of dynamical quark effects will be the main subject of investigation
in the coming years. The comparison of lattice data with effective
low-energy theories such as ChPT has turned into a major
activity. Here, the goal is not just to verify ChPT but to exploit its
predictions in order to make contact with the regime of very light
quarks, which is difficult to access directly in simulations. So far
it is not entirely clear whether the sea quark masses used in
simulations are small enough to justify extrapolations to the chiral
regime based on ChPT at NLO. Simulations of partially quenched QCD
show that the scenario of a massless up-quark is not reflected at all
in the valence quark mass behaviour of pseudoscalar
quantities. Further simulations with smaller quark masses should be
performed to corroborate these findings and to settle the issue of the
mass range in which ChPT at NLO can be expected to be valid.

\vspace{0.1cm}
\par\noindent
{\bf Acknowledgements} It is a pleasure to thank C. Bernard, T. Blum,
C. Dawson, T. Draper, A. Duncan, E. Eichten, S. Hashimoto,
S. Hauswirth, D. Hepburn, J. Hein, K. Holland, T. Kaneko, F. Knechtli,
K.-F. Liu, V. Lubicz, N. Shoresh, R. Sommer, and H. Thacker for
sending their results before the conference and for valuable
discussions about their work.

\end{document}